%% file: main.tex
\documentclass[lettersize,journal]{IEEEtran}
\usepackage{amsmath,amsfonts}
\usepackage{algorithmic}
\usepackage{algorithm}
\usepackage{array}
\usepackage{tabularx}
\usepackage[caption=false,font=normalsize,labelfont=sf,textfont=sf]{subfig}
\usepackage{textcomp}
\usepackage{stfloats}
\usepackage{url}
\usepackage{verbatim}
\usepackage{graphicx}
\usepackage{tikz}
\usetikzlibrary{arrows.meta,positioning,shapes.geometric,fit,calc}
\usepackage{cite}
\makeatletter
\def\@IEEEpubidpullup{2\baselineskip}
\makeatother

\begin{document}
\raggedbottom

\title{From Liability to Asset: A Three-Mode Grid-Forming Control Framework for Centralized Data Center UPS Systems}

\author{Mohamed Shamseldein, Senior Member, IEEE% 
\thanks{M. Shamseldein is an Assistant Professor with the Department of Electrical
	Power and Machines, Faculty of Engineering, Ain Shams University, Cairo,
	Egypt.}}

\IEEEpubid{\parbox{\textwidth}{\centering\scriptsize\setlength{\baselineskip}{8pt}\selectfont This work has been submitted to the IEEE for possible publication.\\Copyright may be transferred without notice, after which this version may no longer be accessible.}}

\maketitle

\input{sections/00_abstract}
\input{sections/00_nomenclature}
\input{sections/01_introduction}
\input{sections/02_background}
\input{sections/03_proposed_framework}
\input{sections/04_simulation}
\input{sections/05_conclusion}
\input{sections/05b_ai_disclaimer}

\bibliographystyle{IEEEtran}
\bibliography{refs}

\end{document}

%% file: sections/00_abstract.tex
\begingroup
% Prevent title-page column balancing from stretching the Abstract--Index Terms gap.
\setlength{\parskip}{0pt}
\begin{abstract}
AI workloads are turning large data centers into highly dynamic power-electronic loads; fault-time behavior and workload pulsing can stress weak-grid points of interconnection. This paper proposes a centralized medium-voltage (MV) uninterruptible power supply (UPS) control architecture implemented as three operating modes: Mode~1 regulates a DC stiff bus and shapes normal-operation grid draw, Mode~2 enforces current-limited fault-mode P--Q priority with UPS battery energy storage system (UPS-BESS) buffering and a rate-limited post-fault ``soft return,'' and Mode~3 optionally provides droop-based fast frequency response via grid-draw modulation. Fundamental-frequency averaged $dq$ simulations (50~MW block, short-circuit ratio (SCR)~=~1.5, 0.5~p.u.\ three-phase dip for 150~ms) show zero unserved information-technology (IT) energy (0.00000~MWh vs.\ 0.00208~MWh for a momentary-cessation benchmark), a 0.57~p.u.\ peak inverter current (vs.\ 1.02~p.u.\ for a synchronous-reference-frame phase-locked loop (SRF-PLL) low-voltage ride-through (LVRT) baseline), a nonzero mean fault-window grid draw of 0.20~p.u.\ (vs.\ $\approx$0 for momentary cessation), and an improved settled point-of-common-coupling (PCC) voltage minimum of 0.79~p.u.\ after one cycle (vs.\ 0.66~p.u.). A forced-oscillation case study applies a 1~Hz pulsed load ($\pm 0.25$~p.u.) and shows that the normal-operation shaping filters the oscillation seen by the grid while the UPS-BESS buffers the pulsing component.
\end{abstract}

\endgroup

%% file: sections/00_nomenclature.tex
\section*{Nomenclature}
\vspace{-0.4em}
\begingroup
\footnotesize
\begin{IEEEdescription}[%
  \settowidth{\labelwidth}{$P_{FFR,max}^{sys}$}%
  \setlength{\labelsep}{0.55em}%
  \setlength{\itemsep}{0.12em}%
  \setlength{\parsep}{0pt}%
  \setlength{\topsep}{0.1em}%
]
\item[$P_{grid}$] Grid-side active power (positive inverter$\rightarrow$grid).
\item[$P_{draw}$] Grid draw ($P_{draw}=-P_{grid}$, positive consumption).
\item[$P_{load}$] IT load (per-unit on 50 MW base).
\item[$P_{bess}$] UPS-BESS active power (positive discharge to IT bus).
\item[$V_{pcc}$] PCC voltage phasor; $|V_{pcc}|$ is its magnitude.
\item[$V_{dc}$] DC-link energy proxy (per-unit voltage-like state).
\item[$I_{max}$] Inverter current limit (per-unit).
\item[$i_{\parallel}/i_{\perp}$] Current components aligned with / orthogonal to $V_{pcc}$ (controls $P$ / $Q$).
\item[$K_v$] Voltage-support gain mapping $(1-|V_{pcc}|)\rightarrow i_{\perp}$.
\item[$\tau_{grid}$] Grid power shaping (LPF) time constant.
\item[$R_{limit}$] Post-fault ramp limit on $P_{draw,ref}$.
\item[$\tau_{bess}, R_{bess}$] UPS-BESS first-order lag time constant and ramp limit.
\item[$t_{det}$] Fault detection/commutation delay before Mode~2 engages.
\item[$\Delta t_{hold}$] DC-side hold-up timing budget for DC-link energy.
\item[$SoC$] UPS-BESS state of charge (proxy), $[SoC_{min},SoC_{max}]$.
\item[$P_{draw,min}^{fault}$] Minimum fault-window grid draw policy (subject to feasibility).
\item[$f,\,f_0$] Bulk-system frequency and nominal frequency (60 Hz).
\item[$x$] Per-unit frequency deviation, $x=(f-f_0)/f_0$.
\item[$K_f$] Mode~3 droop gain mapping $x \rightarrow P_{supp}^{sys}$.
\item[$P_{FFR,max}^{sys}$] Mode~3 maximum fast frequency response (system base).
\end{IEEEdescription}
\endgroup
\vspace{0.2em}

%% file: sections/01_introduction.tex
\IEEEpubidadjcol
\section{Introduction}

\begin{sloppypar}
\IEEEPARstart{T}{his} paper addresses emerging stability risks from large AI data centers connected at weak-grid points of interconnection. These facilities behave as inverter-dominated power-electronic loads with fast power steps, pulsed profiles, and voltage-sensitive fault-time behavior. The July~2025 North American Electric Reliability Corporation (NERC) White Paper on emerging large loads highlights two acute risks \cite{nerc2025}: (i) voltage-sensitive load reduction during transmission faults can produce sudden aggregate load drops (e.g., a $\sim$1.5~GW event in the Eastern Interconnection), and (ii) periodic workload pulsing can excite forced oscillations and reduce damping. These behaviors motivate control strategies that avoid uncontrolled power withdrawal during sags and shape post-fault recovery dynamics.

Centralized MV-UPS architectures provide a single, high-bandwidth control point backed by a consolidated UPS-BESS. Building on this capability, we propose a three-mode supervisory framework: Mode~1 prioritizes internal reliability (a stiff critical bus) and shapes normal-operation grid draw, Mode~2 mitigates adverse grid impacts in weak grids via current limiting, fault-mode P--Q priority with UPS-BESS buffering, and a rate-limited post-fault ``soft return,'' and Mode~3 optionally provides active grid support via controlled load modulation within inverter and UPS-BESS constraints. Mode~2 preempts Modes~1/3 during fault/low-voltage operation.

The novelty of this work is summarized as follows:
\begin{enumerate}
    \item A three-mode supervisory framework that formalizes how a centralized MV-UPS prioritizes internal continuity, adverse-impact mitigation, and optional grid support under inverter and storage constraints.
    \item A current-limited fault-mode allocation strategy (P--Q priority) that retains feasible active power within remaining current headroom while prioritizing reactive support during voltage sags.
    \item A post-fault recovery mechanism (``soft return'') that rate-limits grid power restoration using the UPS-BESS buffer to reduce secondary voltage dips and recovery transients.
    \item A fundamental-frequency averaged $dq$ validation focused on Mode~2, with Mode~1 implemented as a DC stiff-bus regulator and Mode~3 implemented as an optional droop/fast frequency response (FFR) support mode.
\end{enumerate}

The remainder of the paper is organized as follows. Section~\ref{sec:background} reviews relevant architectures and control concepts. Section~\ref{sec:framework} presents the proposed three-mode supervisory framework and the Mode~2 fault-resilience logic. Section~\ref{sec:simulation} reports averaged-$dq$ simulation results, including stress tests and robustness sweeps. Section~\ref{sec:conclusion} concludes and outlines future work.
\end{sloppypar}

%% file: sections/02_background.tex
\section{Background}
\label{sec:background}

This section briefly reviews data center power architectures, grid-forming (GFM) control concepts, and why weak-grid operation motivates coordinated MV-UPS control. We use the PCC to denote the facility point of interconnection (POI).

\subsection{Evolving Data Center Power Architectures}
Distributed UPS architectures embed many small battery-converter units near server racks/rows for modularity and partial efficiency gains \cite{eaton2023, pratt2007}. As AI workloads push power density and facility capacity upward \cite{iea2024}, the distributed approach can limit coordinated grid-facing behavior due to fragmented storage/control and complex aggregate dynamics \cite{zhang2013, dayarathna2016}. Centralized MV-UPS architectures instead consolidate buffering behind a single high-bandwidth control interface, enabling facility-level coordination of fault response and recovery at the point of interconnection.

\subsection{Grid-Forming Control Concepts and Scope}
Grid-following (GFL) inverters regulate current with respect to a measured grid angle, typically obtained using a phase-locked loop (PLL). A representative synchronous-reference-frame PLL is:
\begin{equation}
\dot{\theta}_{pll} = \omega_0 + K_p v_q + K_i \int v_q \, dt
\end{equation}
In weak grids, large phase and voltage excursions can stress PLL tracking and degrade stability \cite{lin2024}.

GFM inverters instead regulate voltage and frequency directly (i.e., behave as controlled voltage sources behind an impedance), often using swing-equation-like outer-loop dynamics such as:
\begin{equation}
J \frac{d\omega}{dt} = P_{ref} - P_{meas} - D(\omega - \omega_{g})
\end{equation}
where $\omega_g$ denotes an external estimate (e.g., from supervisory measurements). In this paper, the validation emphasis is on fault-time current limiting, power buffering, and post-fault recovery shaping in an averaged $dq$ model; detailed angle-generation and frequency-estimation dynamics are left to future work.

For clarity, ``grid-forming'' in this paper refers to a voltage-source control philosophy that remains well-defined under explicit current limiting and prioritizes reactive current during voltage sags without relying on PLL angle tracking for fault response. The presented validation uses a fundamental-frequency averaged $dq$ model with an inner current regulator; detailed angle-generation dynamics (e.g., full virtual synchronous machine (VSM) implementations or virtual oscillator control), switching harmonics, and protection coordination are outside the scope of this paper and are left for future validation. In Sections~III--IV, we describe the implementation using a mode-switching terminology (Mode~1/2/3).

\subsection{Modern GFM Current Limiting and Recovery (Related Work)}
In practical fault conditions, grid-forming controllers must respect converter current and voltage limits. A growing literature addresses how to preserve grid-forming structure under saturation and how to recover smoothly post-fault. Recent approaches include saturation-informed current limiting and fault current limiting schemes that explicitly modify internal references while maintaining a consistent voltage-forming behavior \cite{desai2024saturation, he2024crossforming}, and current-limited dynamics that improve fault-time transient response \cite{boroojeni2024currentlimit}.
The framework proposed here is complementary: it is a supervisory allocation and shaping layer (P--Q priority with minimum fault draw and soft return) that can sit above a chosen grid-forming core provided that the core exposes a current-limited interface for active and reactive objectives. A head-to-head comparison against specific saturation-aware GFM cores is left to electromagnetic transient (EMT)-level validation where switching, inner-loop saturation, and protection interactions can be represented.

\subsection{Related Work for Data-Center Ride-Through and Post-Fault Recovery}
Data-center-specific ride-through and internal voltage management strategies have been proposed, including approaches targeting LVRT of the facility \cite{xie2025dc_lvrt}. At the system level, a broad set of fault-induced delayed voltage recovery (FIDVR) mitigation strategies exist; the ``soft return'' concept here is a facility-side implementation that leverages on-site energy buffering to reduce post-fault load snap-back, and is complementary to grid-side or system-level mitigation schemes \cite{du2021fidvr}.

\subsection{Stability Challenges in Weak Grids}
The integration of gigawatt-scale data centers creates localized "weak grid" conditions, quantitatively defined by the short-circuit ratio, $SCR$:
\begin{equation}
SCR = \frac{S_{sc}}{P_{nom}}
\end{equation}
where $S_{sc}$ is the grid's short-circuit capacity at the Point of Interconnection (POI) and $P_{nom}$ is the data center's total rated load.
As noted in \cite{nerc2025}, large inverter-based loads in low-SCR areas (SCR $< 3$) are prone to:
\begin{enumerate}
    \item \textbf{Voltage sensitivity:} small current changes cause disproportionately large voltage deviations.
    \item \textbf{Control interactions:} fast inverter controls can interact with nearby devices and other large loads.
    \item \textbf{Load withdrawal:} voltage dips can trigger momentary cessation and rapid post-fault ``snap-back''.
\end{enumerate}

IEEE Standard 2800-2022 \cite{ieee2800} addresses these risks by mandating ride-through capabilities and reactive power support for inverter-based resources connected to the transmission system. However, meeting these requirements with standard GFL controls is increasingly difficult in weak grids. The proposed three-mode GFM framework directly addresses these stability gaps by leveraging the inherent robustness of the voltage-source control topology.

%% file: sections/03_proposed_framework.tex
\section{Proposed GFM Control Framework}
\label{sec:framework}

We propose a \emph{grid-forming} supervisory layer for centralized MV-UPS inverters that reallocates active and reactive current objectives during grid disturbances. The framework is implemented as distinct operating control modes: \textit{Mode~1} (Stiff-Bus Mode), \textit{Mode~2} (Fault-Resilience Mode), and \textit{Mode~3} (Grid-Support Mode). Mode~2 preempts Modes~1/3 during fault/low-voltage operation. In the studied scenarios, this meets IEEE~2800 ride-through requirements and exceeds them \cite{ieee2800}.

\begin{figure*}[t]
\centering
\resizebox{\linewidth}{!}{%
\begin{tikzpicture}[font=\footnotesize, >=Latex, node distance=4mm]
  \tikzset{
    blk/.style={draw, rounded corners=2pt, align=center, minimum height=8.2mm, inner sep=3pt},
    plant/.style={blk, fill=blue!4, draw=blue!55!black},
    ctrl/.style={blk, fill=green!5, draw=green!45!black},
    logic/.style={blk, fill=red!4, draw=red!55!black, dashed},
    aux/.style={blk, fill=gray!6, draw=black!55},
    sw/.style={draw, trapezium, trapezium angle=70, shape border rotate=270, minimum height=15mm, minimum width=10mm, inner sep=3pt, fill=white, label=center:SW},
    pline/.style={-Latex, line width=0.95pt, draw=blue!60!black},
    cline/.style={-Latex, line width=0.85pt, draw=green!45!black},
    fline/.style={-Latex, line width=0.85pt, draw=red!60!black},
    thin/.style={-Latex, line width=0.65pt, draw=black!60},
    annot/.style={font=\footnotesize\itshape, inner sep=1pt},
  }

  % Top (plant/interface)
  \node[plant, minimum width=18mm] (grid) {Grid\\$(V_{th})$};
  \node[plant, minimum width=22mm, right=6mm of grid] (rl) {RL filter\\$(R_f,\,X_f)$};
  \node[plant, minimum width=22mm, right=6mm of rl] (vsc) {UPS inverter\\(avg.)};

  \draw[pline] (vsc.west) -- node[annot, above] {$v_{inv,dq}$} (rl.east);
  \draw[pline] (rl.west) -- node[annot, above] {$i_g$} (grid.east);
  \node[annot] at ($(rl.south)+(0,-2.0mm)$) {$v_{pcc},\,i_g$};
  \draw[thin] (rl.south) -- ++(0,-8mm);

  % Bottom (control)
  \node[ctrl, minimum width=22mm, below=17mm of rl] (meas) {$abc \rightarrow dq$\\transform};
  \node[logic, minimum width=26mm, right=5mm of meas] (faultdet) {Fault detect\\\& logic};

  % Explicit switch/multiplexer
  \node[sw, right=25mm of faultdet] (sw) {};
  
  % Outer loop chain
  \node[ctrl, minimum width=26mm, right=5mm of sw] (outer) {Outer loops\\($P_{draw}$, BESS)};
  \node[ctrl, minimum width=22mm, right=5mm of outer] (rate) {Soft return\\rate limit};
  \node[ctrl, minimum width=24mm, right=5mm of rate] (inner) {Inner current\\PI + clamp};
  \node[aux, minimum width=22mm, right=5mm of inner] (mod) {$dq \rightarrow abc$\\PWM cmd.\\to inverter};

  % Parallel mode-specific setpoint generators (stacked on left)
  \node[ctrl, minimum width=30mm, above=5mm of faultdet] (normal) {Normal setpoints\\Mode~1 + Mode~3\\($P_{load}$, SoC, $f$)};
  \node[logic, minimum width=30mm, below=5mm of faultdet] (faultmode) {Fault setpoints\\Mode~2\\(P--Q priority)};

  \node[aux, minimum width=22mm, below=12mm of grid] (align) {Frame align\\($V_{pcc}$)};

  % Wiring: measurements
  \draw[thin] (rl.south) -- (meas.north);
  \draw[thin] (align.east) .. controls +(+10mm,0mm) and +(-10mm,0mm) .. (meas.west);
  \draw[cline] (meas.east) -- (faultdet.west);

  % Logic triggers fault mode setpoints
  \draw[fline] (faultdet.south) -- (faultmode.north);

  % Sources into MUX Inputs (West side of rotated trapezium)
  \draw[cline] (normal.east) -| node[annot, pos=0.8, above] {0} ([xshift=-2mm]sw.130) -- (sw.130);
  \draw[fline] (faultmode.east) -| node[annot, pos=0.8, below] {1} ([xshift=-2mm]sw.230) -- (sw.230);

  % Control input (Dashed, from Logic)
  \draw[thin, dashed] (faultdet.east) -- node[annot, above] {sel} (sw.west);

  % Switch output
  \draw[cline] (sw.east) -- (outer.west);
  \draw[cline] (outer.east) -- (rate.west);
  \draw[cline] (rate.east) -- (inner.west);
  \draw[cline] (inner.east) -- (mod.west);

  % \node[annot, above=0.5mm of sw] {select path};

  % Actuation is the inverter PWM command (shown in the block label).

  % Feedback loop (conceptual return path)
  \draw[thin] (grid.south) -- (align.north);
\end{tikzpicture}%
}
\caption{High-level block diagram of the averaged control structure implemented in simulation. Mode switching is represented explicitly as a switch/multiplexer: Mode~2 preempts Modes~1/3 during fault/low-voltage conditions by selecting the fault setpoint path (current-limited P--Q priority), while normal operation selects the Mode~1/3 setpoint path (DC stiff bus with optional droop/FFR). The selected setpoints drive a common outer-loop chain with rate-limited recovery, and the UPS-BESS buffers the resulting active-power mismatch within constraints.}
\label{fig:block_diagram}
\end{figure*}
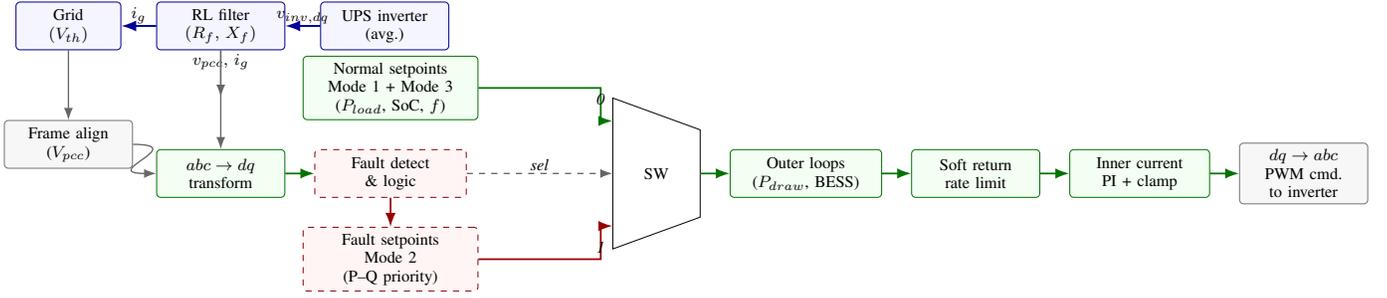

Fig.~\ref{fig:block_diagram} summarizing the averaged control structure. The logic is intentionally expressed in terms of measurable quantities ($|V_{pcc}|$, $i_d$, $i_q$) and explicit saturation operators to make limit behavior reproducible.

\subsection{Mode 1: Internal Reliability (DC Stiff Bus)}
\label{sec:stage1}
Mode~1 targets internal continuity by maintaining a stiff critical bus using the centralized UPS-BESS and the DC-link energy reservoir. In the averaged implementation used in this paper, the internal bus is represented by a DC-link proxy voltage $V_{dc}$ whose energy balance is discussed in Section~\ref{sec:simulation}. During normal (non-fault) operation, Mode~1 regulates $V_{dc}$ toward a reference $V_{dc,ref}$ by adding a bounded stiffness term to the UPS-BESS power command:
\begin{equation}
P_{vdc}=K_{p,dc}(V_{dc,ref}-V_{dc})+K_{i,dc}\xi_{dc}, \quad \dot{\xi}_{dc}=V_{dc,ref}-V_{dc},
\end{equation}
with $P_{vdc}$ clipped by UPS-BESS power, ramp-rate, and SoC constraints. A slow SoC restoration bias can also be applied by slightly increasing the grid-draw target in normal operation (charging the UPS-BESS) to return SoC toward a desired operating point without creating a secondary disturbance.

\textbf{Coordination with Mode~2:} Mode~2 has priority. Mode~1 regulation is disabled during fault/low-voltage operation to avoid conflicting objectives.

\subsection{Mode 2: Fault Resilience via Active/Reactive Power Decoupling}
Standard ``constant-power'' loads (e.g., server power supplies) are detrimental during voltage sags because they attempt to maintain $P$ as $V$ falls ($P = V \downarrow \times I \uparrow$), often triggering overcurrent protection and resulting in simultaneous disconnection.

The proposed controller implements a ``P--Q split'' strategy that activates when the PCC voltage magnitude falls below a threshold, $|V_{pcc}| < V_{thresh}$ (typically $0.85$~p.u.).

\subsubsection{Active Power ($P$) Decoupling}
Upon detection of a fault ($|V_{pcc}| < V_{thresh}$), the controller adjusts the grid-side active current reference, $i_d^*$, to prioritize stability while avoiding destabilizing constant-power behavior. Rather than forcing $i_d^*=0$, the proposed logic retains as much active current as the inverter current limit allows after allocating reactive current for voltage support. This results in a controlled reduction of $P_{grid}$ during the dip, bounded by inverter ampacity and UPS-BESS capability.

\noindent This avoids uncontrolled momentary cessation and the ``$P=VI$'' current amplification associated with constant-power behavior in low-voltage conditions, while preserving thermal headroom by operating within a strict current limit.

\subsubsection{Reactive Power ($Q$) Injection}
With active current reduced as needed, the remaining current headroom under the inverter limit $I_{max}$ becomes available for reactive current injection ($i_q$). The GFM controller acts as a voltage source, injecting capacitive current to support the PCC voltage:

\begin{equation}
i_q^* = \min \left( I_{max}, \ K_v (V_{ref} - |V_{pcc}|) \right)
\end{equation}

With reactive priority, the available active current headroom is:
\begin{equation}
|i_d^*| \le \sqrt{I_{max}^2 - (i_q^*)^2}
\end{equation}

\noindent The data center transforms from a load into a dynamic reactive-power support device (static synchronous compensator (STATCOM)-like behavior) during the fault, aiding protection clearing and neighbor load ride-through.

\subsubsection{Managing the $P$--$Q$ Trade-Off at Scale}
When many large inverter-interfaced loads respond simultaneously, a hard reduction of active power draw can manifest as a bulk-system load-drop event. To provide an explicit control lever between voltage support and aggregate active-power withdrawal, the proposed implementation reserves a minimum active-power draw during faults, $P_{draw,min}^{fault}$ (subject to current limits):
\begin{equation}
P_{draw} \ge P_{draw,min}^{fault}
\end{equation}
This policy reduces the maximum achievable reactive current during deep dips, but it mitigates the risk of simultaneous near-zero active power draw across large load blocks. The remaining deficit power is supplied by the UPS-BESS subject to its constraints.

\subsubsection{Priority and Saturation Pseudocode}
The fault-mode current references use reactive priority with a vector current limit and an explicit minimum active-power draw policy. Notation: $P_{grid}$ is positive from inverter to grid and $P_{draw}=-P_{grid}$ is positive when the data center consumes grid power. In the simulation, $P$ and $Q$ are allocated in a $V_{pcc}$-aligned basis (so $i_\parallel$ controls $P$ and $i_\perp$ controls $Q$ even when $v_q\neq 0$), then mapped back to $dq$ currents for the inner PI loop.
In this $V_{pcc}$-aligned basis, the ``$d$-axis'' voltage used for $P$--$I$ conversion is approximated by $v_\parallel \approx |V_{pcc}|$.

\begin{algorithm}[htbp]
\caption{Fault-Mode P--Q Priority with UPS-BESS Buffering}
\begin{algorithmic}[1]
\STATE Measure $|V_{pcc}|$, compute $inFault \leftarrow (|V_{pcc}| < V_{thresh})$ after detection delay $t_{det}$
\STATE $i_{\perp,des} \leftarrow \mathrm{sat}\!\left(K_v(V_{ref}-|V_{pcc}|),\,0,\,I_{max}\right)$
\STATE $P_{draw,min} \leftarrow \max(0, P_{load}-P_{bess,dis}^{max})$
\IF{$inFault$}
\STATE $P_{draw,min} \leftarrow \max(P_{draw,min}, P_{draw,min}^{fault})$
\ENDIF
\STATE $i_{\parallel,req} \leftarrow P_{draw,min}/(1.5\,|V_{pcc}|)$
\STATE $i_\perp^* \leftarrow -\min(i_{\perp,des}, \sqrt{I_{max}^2-i_{\parallel,req}^2})$
\STATE $|i_\parallel|_{max} \leftarrow \sqrt{I_{max}^2-(i_\perp^*)^2}$
\STATE Shape grid draw reference with LPF + rate limits (Eq.~(\ref{eq:pdraw_shape})): $P_{draw,ref} \leftarrow \mathrm{shape}(P_{draw,target})$
\STATE $P_{bess,cmd} \leftarrow \mathrm{sat}(P_{load}-P_{draw,ref}, -P_{bess,chg}^{max}, P_{bess,dis}^{max})$
\STATE $P_{draw,cmd} \leftarrow \max(0, P_{load}-P_{bess})$
\STATE $i_\parallel^* \leftarrow \mathrm{sat}\!\left(-P_{draw,cmd}/(1.5\,|V_{pcc}|),\, -|i_\parallel|_{max},\, |i_\parallel|_{max}\right)$
\STATE Map $(i_\parallel^*, i_\perp^*)$ to $(i_d^*, i_q^*)$ and track with PI current control (integrator clamped) and $|v_{inv}|\le E_{max}$
\end{algorithmic}
\end{algorithm}
where $\mathrm{sat}(\cdot)$ is defined in Section~\ref{sec:simulation} and $\mathrm{shape}(\cdot)$ denotes the rate-limited first-order filter in Eq.~(\ref{eq:pdraw_shape}).

\subsubsection{UPS-BESS Power Balance and Limits}
The centralized DC energy reservoir supplies the difference between the IT load and the grid-side active power during transients:
\begin{equation}
P_{bess} = P_{load} - P_{draw}
\end{equation}
subject to practical constraints on instantaneous power, ramp rate, and energy:
\begin{align}
|P_{bess}| &\le P_{bess,max}, \quad \left|\frac{dP_{bess}}{dt}\right| \le R_{bess} \\
SoC_{min} &\le SoC \le SoC_{max}
\end{align}

\subsubsection{Practical Considerations: DC-Side Bandwidth and Thermal Limits}
The simulations model the UPS-BESS power path using an explicit first-order lag and ramp limits (Section~\ref{sec:simulation}), capturing finite DC/DC power-path bandwidth at a control-oriented level. In hardware, the DC/DC converter and DC bus capacitance must sustain the IT load during the interval from fault inception to \emph{meaningful} UPS-BESS power transfer (measurement, detection logic, gating/commutation, and converter pickup). A practical timing budget is
\begin{equation}
\Delta t_{hold} \approx t_{meas}+t_{logic}+t_{gate}+t_{dc/dc},
\end{equation}
where a conservative pickup time for a first-order, ramp-limited power path is $t_{dc/dc}\ge \max(3\tau_{bess}, \Delta P/R_{bess})$ for a step of size $\Delta P$.
A simple DC hold-up energy relationship is
\begin{equation}
\frac{1}{2}C_{dc}\left(V_{dc,0}^2 - V_{dc,min}^2\right) \ge P_{load}\,\Delta t_{hold},
\end{equation}
In the per-unit DC-link energy proxy used in Section~\ref{sec:simulation}, this bound maps directly to a minimum required energy constant via $\Delta t_{hold}\le (V_{dc,0}^2-V_{dc,min}^2)T_{dc}/(2P_{load})$ under worst-case loss of input power.
Accordingly, Section~\ref{sec:simulation} explicitly includes a fault-detection delay before Mode~2 engages and evaluates stress cases that increase $t_{meas}+t_{logic}$ or reduce DC-side bandwidth/energy to illustrate graceful degradation when these constraints bind.
As an order-of-magnitude example, a 50~MW block with $\Delta t_{hold}=10$~ms requires 0.5~MJ of hold-up energy; for a 10~kV-class DC link and $V_{dc,min}=0.7V_{dc,0}$, this corresponds to $C_{dc}\approx 20$~mF.

In addition, although reactive-priority operation holds the current magnitude within $I_{max}$, the rapid shift in current phase angle can affect semiconductor and passive-component thermal cycling; practical implementations should incorporate duration limits, thermal derating, and/or adaptive reactive support under sustained low-voltage conditions.

\subsubsection{Rate-Limited Recovery (Post-Fault Transition)}
A critical vulnerability identified in NERC disturbance reports is FIDVR, often exacerbated when large loads snap back to full power the instant voltage recovers.

To mitigate this, the controller uses the UPS-BESS buffer to implement a rate-limited ``soft return'' of grid draw.
Upon voltage recovery, defined as $|V_{pcc}| > V_{rec}$ (e.g., $0.9$~p.u.), the controller ramps the grid-draw reference $P_{draw,ref}$ back toward nominal rather than applying an immediate step change, enforcing a rate limit to mitigate secondary transients:

\begin{equation}
\left| \frac{d P_{draw,ref}}{dt} \right| \le R_{limit} \quad (e.g.,\ 10~MW/s)
\end{equation}

The UPS-BESS supplies the deficit power ($P_{load} - P_{draw,ref}$) throughout this transition, provided the State of Charge (SoC) remains above a critical level ($SoC_{min}$). This seamless handover helps prevent secondary voltage dips and inrush-like transients, enabling a grid-friendly restoration that is difficult to achieve with uncoordinated distributed server power supplies.

\subsection{Mode 3 (Optional): Droop-Based Fast Frequency Response via Grid-Draw Modulation}
\label{sec:stage3}
Mode~3 leverages the same centralized UPS-BESS flexibility used in Mode~2 to provide \emph{optional} bulk-system support during frequency events in normal (non-fault) operation. Rather than exporting power, the UPS acts as a controllable load that can temporarily \emph{reduce} (under-frequency) or \emph{increase} (over-frequency) its grid draw while maintaining the IT load by discharging/charging the UPS-BESS within its constraints.

Let $f$ be an available frequency estimate (e.g., from a phasor measurement unit (PMU), plant-wide measurement, or a synchronization block) and define the per-unit frequency deviation $x=(f-f_0)/f_0$. A simple droop/FFR law on an external system base is
\begin{equation}
P_{supp}^{sys}=\mathrm{sat}\!\left(K_f(-x),\, -P_{FFR,max}^{sys},\, P_{FFR,max}^{sys}\right),
\end{equation}
where $P_{supp}^{sys}>0$ corresponds to \emph{load relief}. The command is mapped to the data-center base as $P_{supp}=P_{supp}^{sys}\,S_{sys}/(N\,P_{nom})$ and applied as a modification of the grid-draw target:
\begin{equation}
P_{draw,target}= \mathrm{sat}\!\left(P_{load}-P_{supp},\, 0,\, P_{load}+P_{chg}^{max}\right).
\end{equation}
The UPS-BESS supplies the resulting mismatch $P_{bess}=P_{load}-P_{draw}$ subject to power, ramp, and SoC limits (Section~\ref{sec:simulation}).

\textbf{Coordination with Mode~2:} Mode~2 has priority. Mode~3 is enabled only when the PCC is outside the low-voltage region (e.g., $|V_{pcc}|>V_{thresh}$) and is disabled during fault/ride-through to avoid conflicting objectives between voltage support (reactive priority) and bulk-frequency support (active modulation).
In implementation, the commanded modulation is also clipped by inverter and storage feasibility (UPS-BESS power/SoC limits and the current-limited feasible draw range).

%% file: sections/04_simulation.tex
\section{Simulation Results}
\label{sec:simulation}

We validate the proposed three-mode framework using fundamental-frequency time-domain simulations of a nonlinear averaged $dq$ model. Results are organized by operating mode: Mode~1 (normal-operation functions), Mode~2 (fault ride-through and grid-friendly recovery), and Mode~3 (optional frequency support via grid-draw modulation).

\subsection{Methodology \& System Parameters}
We simulate a 50~MW data center block (1.0~p.u.) at 13.8~kV connected to a Thevenin grid equivalent through an $RL$ filter ($X_f=0.15$~p.u.). The grid is ultra-weak (SCR~=~1.5, X/R~=~5), consistent with the NERC risk context \cite{nerc2025}. The IT load steps to 1.0~p.u.\ at $t=0.1$~s. A balanced three-phase voltage dip to 0.5~p.u.\ is applied from $t=0.50$~s to $t=0.65$~s.

In all cases the inverter current is limited to $I_{max}=1.0$~p.u.\ (unit ampacity). The UPS-BESS is modeled as a limited power buffer (power/ramp/SoC constraints) that supplies the mismatch between the IT load and the commanded grid draw. Sign convention: $P_{grid}>0$ is inverter-to-grid, and $P_{draw}=-P_{grid}>0$ is grid draw.

All cases share the same inner current-regulator gains and inverter voltage/current limits; the GFL benchmark tuning (PLL/LVRT parameters) is listed below. The supervisory logic is expressed with explicit saturations and first-order buffer dynamics so that limit behavior is reproducible.

\subsubsection{Explicit Averaged Equations (Controller and Limits)}
We use $\mathrm{sat}(x,a,b)=\min(\max(x,a),b)$. The grid-draw reference is shaped by a first-order filter with asymmetric rate limits:
\begin{equation}
\begin{split}
\dot{P}_{draw,ref} &= \mathrm{sat}\!\left(\frac{P_{draw,target}-P_{draw,ref}}{\tau_{grid}},\right.\\
&\left.\qquad -R_{down},\ R_{limit}\right).
\end{split}
\label{eq:pdraw_shape}
\end{equation}
The UPS-BESS power is modeled as a first-order lag with a ramp limit and SoC proxy:
\begin{align}
\frac{dP_{bess}}{dt} &= \mathrm{sat}\!\left(\frac{P_{bess,cmd}-P_{bess}}{\tau_{bess}},\ -R_{bess},\ R_{bess}\right) \\
\frac{dSoC}{dt} &= -\frac{P_{bess}}{T_{autonomy}}
\end{align}
where $P_{bess,cmd}=\mathrm{sat}(P_{load}-P_{draw,ref},-P_{chg}^{max},P_{dis}^{max})$ with additional SoC gating. Current control uses PI regulators on $i_d$ and $i_q$ with integrator clamping (anti-windup proxy) and a voltage magnitude limit $|v_{inv}|\le E_{max}$.
\noindent The PI implementation is:
\begin{equation}
v_{pi}=K_p(i^*-i)+K_i\,\mathrm{sat}(\xi,-\xi_{lim},\xi_{lim}), \quad \dot{\xi}=i^*-i.
\end{equation}

To implement Mode~1 (Stiff-Bus Mode) in the averaged model, the internal DC-link is modeled as an energy state and regulated toward a reference $V_{dc,ref}$:
\begin{equation}
\frac{d}{dt}\left(\frac{1}{2}V_{dc}^2\right) = \frac{P_{in}-P_{load}}{T_{dc}}, \quad P_{in}=P_{draw}+P_{bess}
\end{equation}
and the inverter voltage capability is bounded by the DC-link voltage as $|v_{inv}|\le E_{max}V_{dc}$.
Here, $V_{dc}$ is an energy proxy (not a switched DC-link model). In normal operation, a bounded DC-bus regulation term $P_{vdc}$ (Section~\ref{sec:stage1}) is added to the UPS-BESS power command; it is disabled during Mode~2. Mode~2 engages after a fixed detection/commutation delay $t_{det}$, and pickup is limited by $\tau_{bess}$ and $R_{bess}$. Under worst-case loss of input power ($P_{in}\!\approx\!0$), the proxy gives $t_{hold}\approx (V_{dc,0}^2-V_{dc,min}^2)T_{dc}/(2P_{load})$ to reach $V_{dc,min}$.

\subsubsection{Limitations and Graceful Degradation Tests}
The model is averaged (no PWM harmonics, detailed protection, or full converter saturation), but it includes explicit current/voltage limits, PI anti-windup, UPS-BESS lag/ramp/SoC constraints, detection delay, and a DC-energy proxy. To show graceful degradation when constraints bind, we run stress tests (reduced DC energy, increased delay, reduced UPS-BESS bandwidth/power/ramp) and a ``no BESS'' case; Table~\ref{tab:stress} summarizes fault-window metrics.

In the Thevenin-equivalent network ($V_{pcc}=V_{th}+Z_{th}I$ under our sign convention), reducing fault-time current draw can increase $|V_{pcc}|$ by reducing the drop across $Z_{th}$, at the cost of greater UPS-BESS burden and/or unserved IT energy when buffering limits are reached.

\subsection{Mode 1: Normal-Operation Functions}
\subsubsection{Forced-Oscillation Filtering}
\begin{figure}[htbp]
\centering
\captionsetup{font=footnotesize}
\captionsetup[subfloat]{font=footnotesize}
\subfloat[Grid-draw shaping (ramping/LPF) under a load step.]{%
\includegraphics[width=\columnwidth]{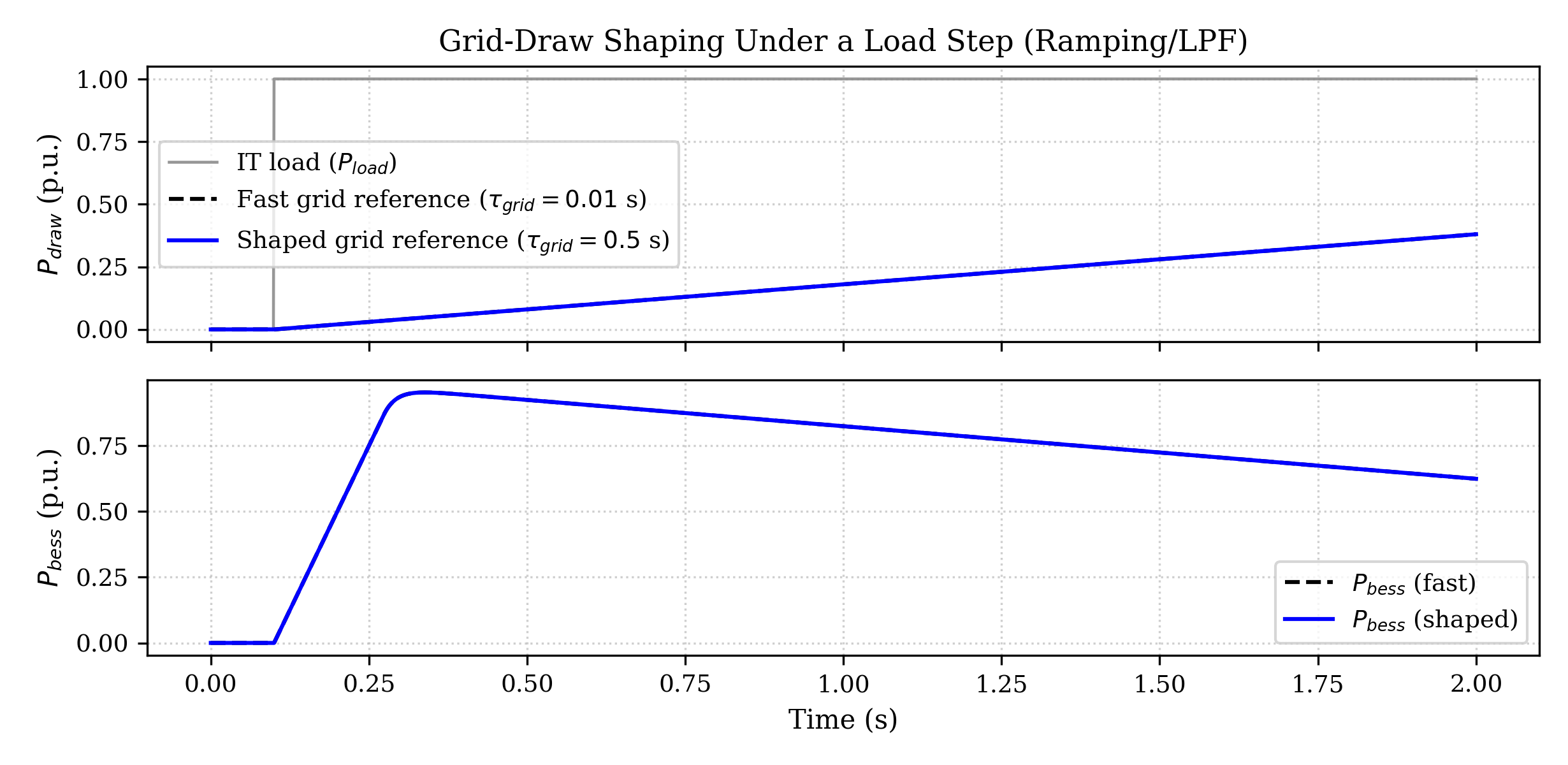}}\\
\subfloat[Forced-oscillation filtering under a 1~Hz pulsed load.]{%
\includegraphics[width=\columnwidth]{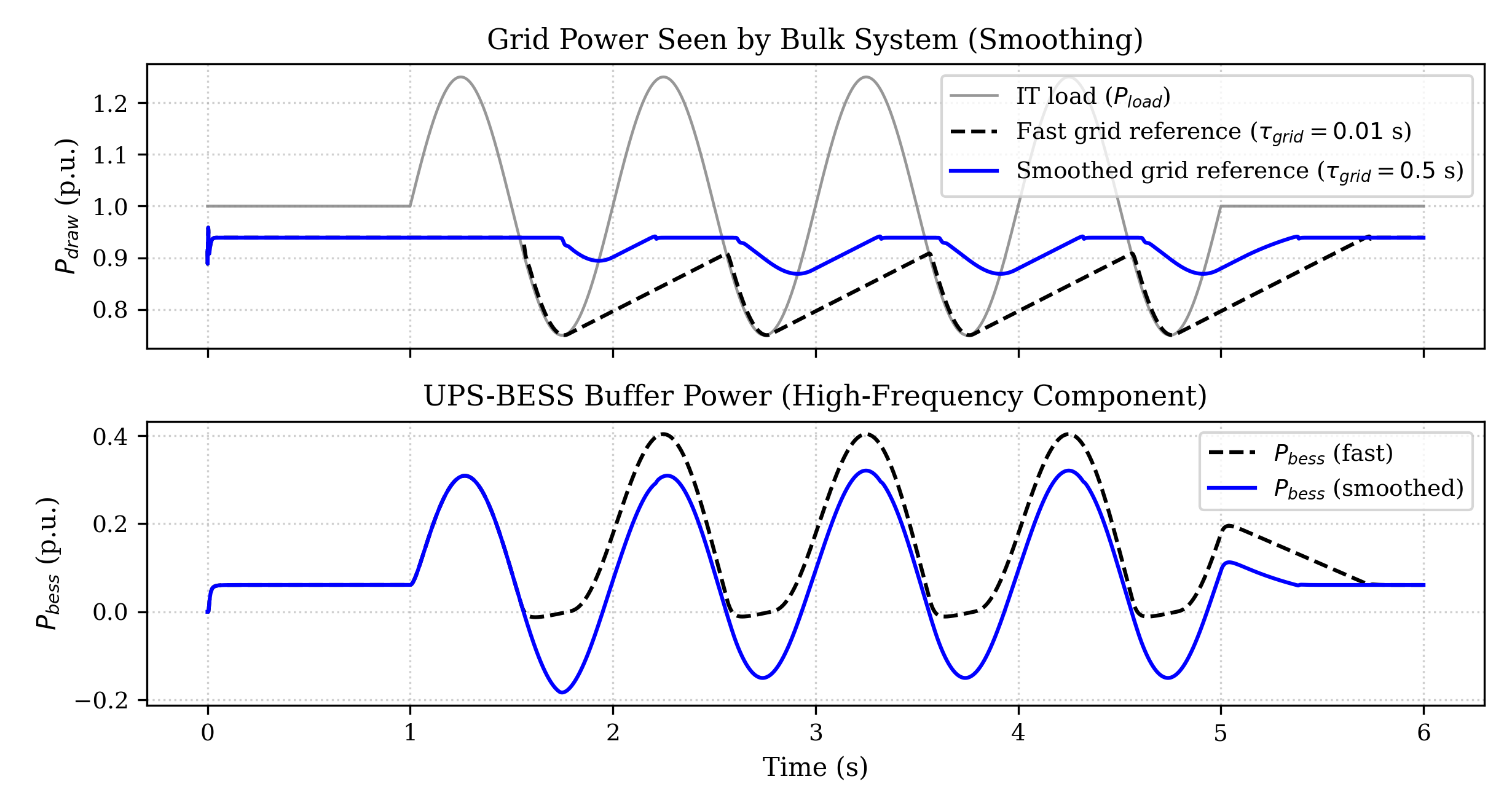}}
\caption{Mode~1 normal-operation power shaping: (a) the ramping behavior is an intentional choice from the grid-draw shaping block (finite $\tau_{grid}$) under a step in IT load, and (b) under a 1~Hz pulsed load the UPS-BESS buffers the high-frequency component, reducing the oscillatory active power seen by the grid.}
\label{fig:pulse_results}
\end{figure}

Mode~1 includes a normal-operation power shaping function that intentionally limits how fast the facility's grid draw follows internal load variations. Fig.~\ref{fig:pulse_results}(a) isolates this behavior under a step in IT load: the grid-draw shaping block (finite $\tau_{grid}$) produces a gradual ramp in $P_{draw}$, while the UPS-BESS supplies the transient mismatch. Fig.~\ref{fig:pulse_results}(b) then demonstrates forced-oscillation filtering under a 1~Hz pulsed load component ($\pm 0.25$~p.u.): the shaping attenuates the 1~Hz component seen by the grid, and the UPS-BESS buffers the oscillatory residual within power/SoC limits.

\subsubsection{DC Stiff-Bus Regulation}
To demonstrate Mode~1 as a closed-loop internal reliability function (not only an energy proxy), we initialize the model with a DC-link energy deficit ($V_{dc}(0)<V_{dc,ref}$) and run a no-fault pulsed-load scenario. With Mode~1 disabled, $V_{dc}$ does not recover in the averaged energy proxy because $P_{in}\approx P_{load}$ on average. With Mode~1 enabled, the DC stiff-bus regulator injects a bounded stiffness term via the UPS-BESS to restore $V_{dc}$ toward $V_{dc,ref}$ while respecting UPS-BESS limits. Mode~2 remains inactive because $|V_{pcc}|>V_{thresh}$.

\begin{figure}[htbp]
\centerline{\includegraphics[width=\columnwidth]{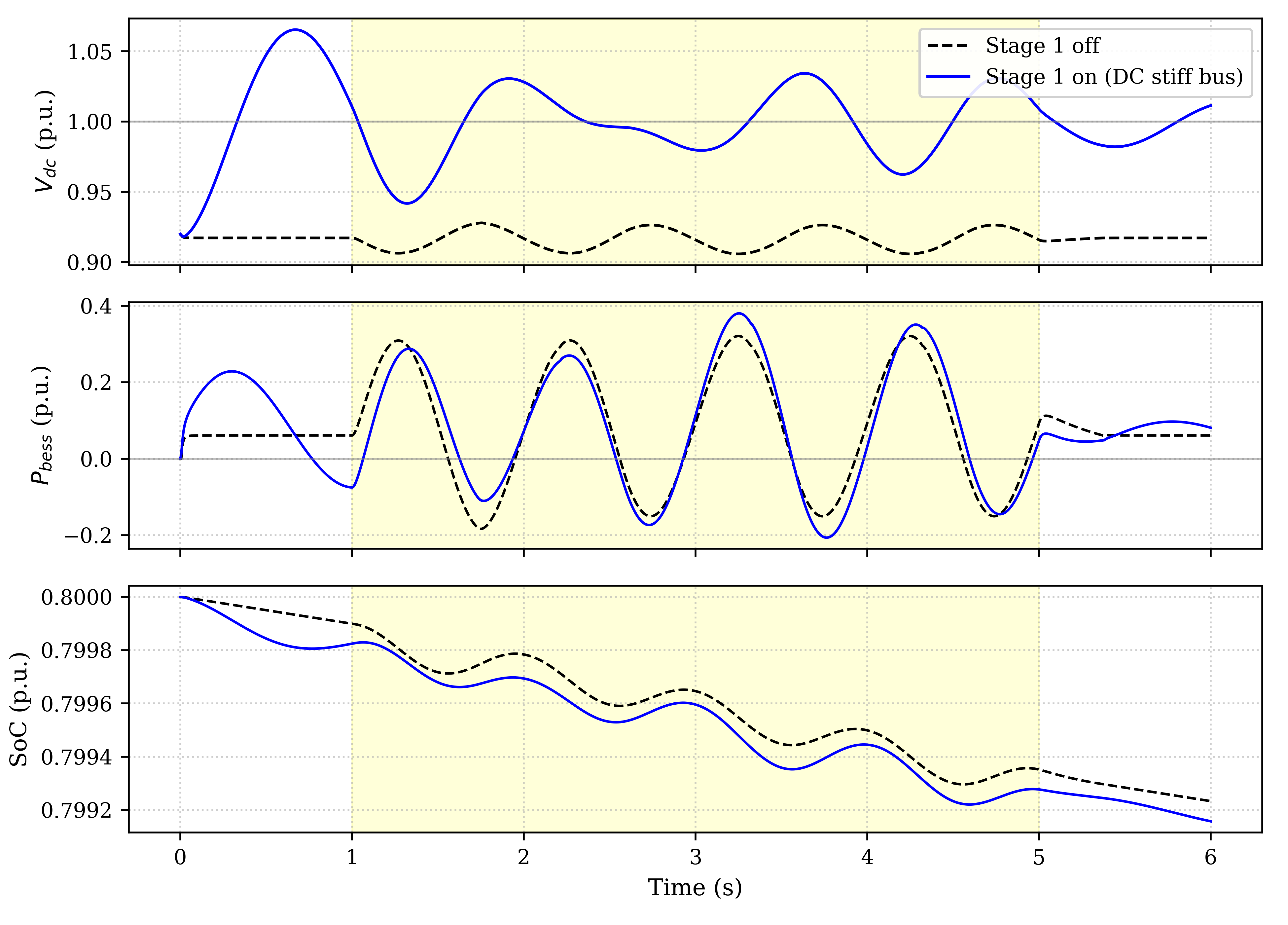}}
\caption{Mode~1 (DC stiff bus) demonstration under a no-fault pulsed-load scenario with an initial DC-link energy deficit ($V_{dc}(0)<V_{dc,ref}$): comparison of $V_{dc}$, $P_{bess}$, and SoC with Mode~1 disabled vs enabled. The shaded window indicates the pulsed-load interval.}
\label{fig:stage1_stiffbus}
\end{figure}

\subsection{Mode 2: Fault Ride-Through and Recovery}
\subsubsection{Weak-Grid Fault Resilience and Baseline Comparison}
The first scenario examines the ride-through response to a severe three-phase voltage dip (retained voltage 0.5~p.u.) lasting 150~ms under an ultra-weak node (SCR~=~1.5).

\paragraph{Benchmark: Standard GFL Control (Momentary Cessation)}
The benchmark represents a conventional grid-following (GFL) implementation: constant-power draw with inner current regulation plus momentary cessation (currents forced to zero when $|V_{pcc}|<0.70$~p.u.), consistent with the voltage-sensitive load reduction behavior highlighted in NERC reports \cite{nerc2025}.

\paragraph{Benchmark: GFL with SRF-PLL and LVRT Reactive Priority}
To provide a stronger grid-following baseline, we also evaluate an SRF-PLL-based GFL controller with low-voltage reactive current priority under a vector current limit (i.e., LVRT-type behavior). This baseline does not invoke momentary cessation and is intended to be closer to an IEEE~2800-style ride-through response \cite{ieee2800}.
The PLL and LVRT parameters are selected to provide stable angle tracking for the SCR~=~1.5 test case while enforcing the same current and voltage limits as the other cases (SRF-PLL gains $K_{p,pll}/K_{i,pll}=20/200$, LVRT threshold $V_{thresh}=0.85$~p.u., LVRT reactive gain $K_v=2.5$).

\begin{figure}[htbp]
\centerline{\includegraphics[width=\columnwidth]{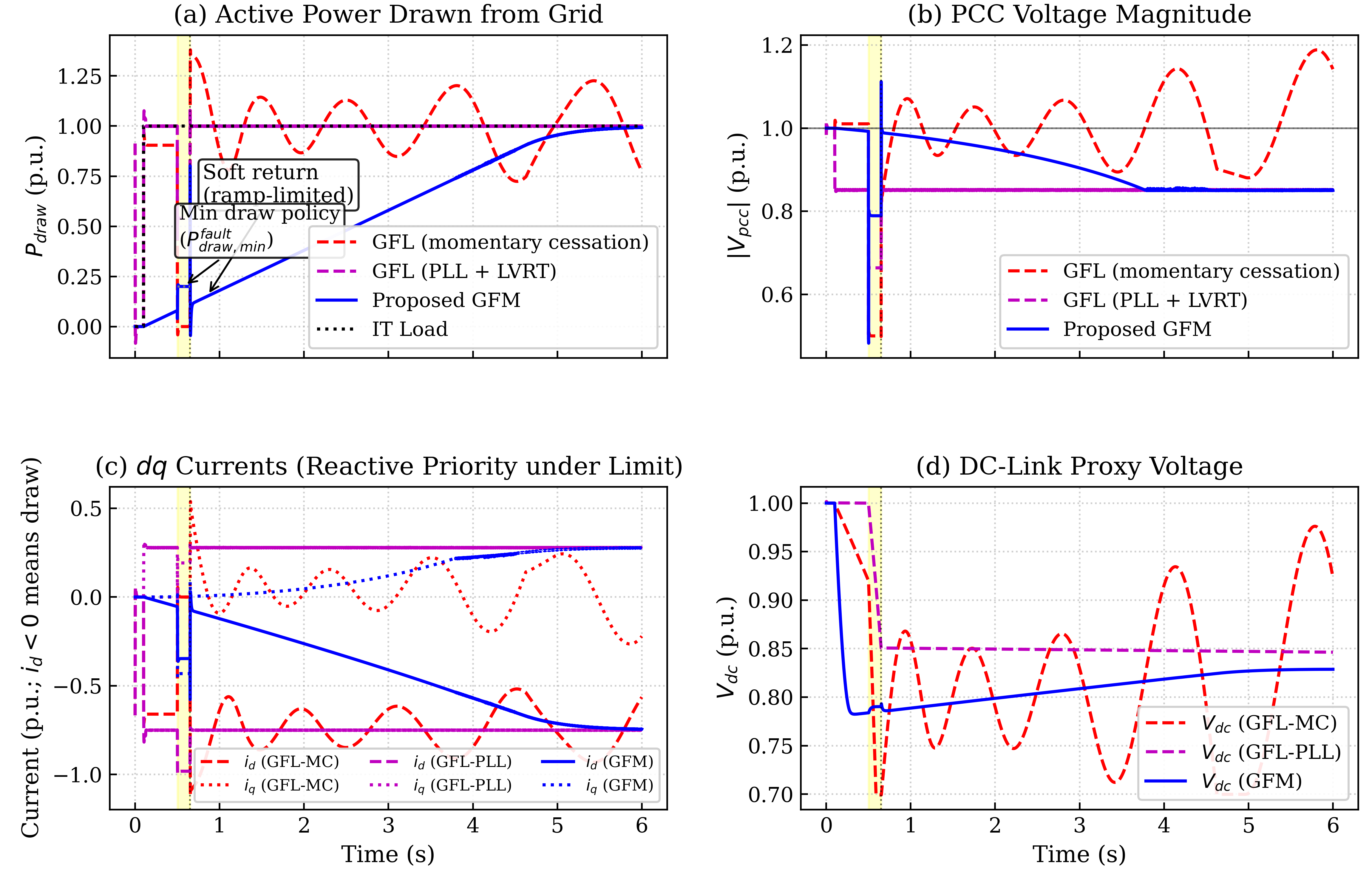}}
\caption{Averaged $dq$ simulation results comparing two GFL baselines to the proposed Mode~2 controller: GFL with momentary cessation (red), GFL with SRF-PLL and LVRT reactive priority (magenta), and proposed controller (blue). The three-phase voltage dip is applied from $t=0.50$~s to $t=0.65$~s (yellow band). Panels show (a) active power drawn from the grid (IT load shown as reference; dotted line indicates the minimum-draw policy $P_{draw,min}^{fault}$), (b) $|V_{pcc}|$, (c) $dq$ currents under a vector current limit (with $i_d<0$ indicating net grid draw under the adopted sign convention), and (d) DC-link proxy voltage $V_{dc}$.}
\label{fig:sim_results}
\end{figure}

\paragraph{Proposed GFM Framework (Mode 2)}
In contrast, the proposed Mode~2 control activates the ``P--Q split'' logic when $|V_{pcc}|<0.85$~p.u.:
\begin{enumerate}
    \item \textbf{Reactive-priority current limiting:} Allocate $i_q^*$ for voltage support, and retain active current within the remaining headroom ($|i_d^*|\le\sqrt{I_{max}^2-(i_q^*)^2}$) to avoid constant-power current amplification without forcing $P\!\rightarrow\!0$.
    \item \textbf{UPS-BESS buffering:} Supply the IT-load/grid-draw mismatch subject to power/ramp/SoC constraints.
\end{enumerate}
This removes constant-power behavior during the dip while maintaining a continuous, controlled response at the PCC. Fig.~\ref{fig:sim_results}(a) also shows the minimum-draw policy ($P_{draw,min}^{fault}$) as a reference; it is enforced when feasible under the current limit and PCC voltage, but can be violated under extreme voltage binding.

\subsubsection{Grid-Friendly Recovery}
This scenario evaluates post-fault recovery to mitigate FIDVR. Instead of an immediate ``snap back'' to full power after clearing, the soft return ramps the grid active-power command to nominal at $10$~MW/s (0.2~p.u./s on a 50~MW base) while the UPS-BESS buffers the mismatch. We report both the instantaneous minimum $|V_{pcc}|$ during the dip and a ``settled'' minimum that excludes the first 60~Hz cycle after fault inception to reduce sensitivity to detection/commutation transients.

\begin{table}[!htbp]
\caption{Comparative Assessment of Fault Response (SCR = 1.5)}
\vspace{-0.6\baselineskip}
\label{tab:comparison}
\centering
\footnotesize
\setlength{\tabcolsep}{3pt}
\renewcommand{\arraystretch}{1.05}
\begin{tabularx}{\columnwidth}{@{}>{\raggedright\arraybackslash}Xccc@{}}
\hline\hline
\textbf{Metric} & \textbf{GFL-MC} & \textbf{GFL-PLL} & \textbf{Proposed} \\
\hline
Momentary cessation (MC) used? & Yes & No & No \\
$\min |V_{pcc}|$ during dip (p.u.) & $0.491$ & $0.503$ & $0.482$ \\
\shortstack{$\min |V_{pcc}|$ during dip\\(after 1 cycle) (p.u.)} & $0.499$ & $0.661$ & $0.789$ \\
$\max |I|$ during dip (p.u.) & $0.658$ & $1.017$ & $0.570$ \\
$\min V_{dc}$ during dip (p.u.) & $0.699$ & $0.851$ & $0.784$ \\
\shortstack{Unserved IT\\energy (MWh)} & $0.00208$ & $0.00096$ & $0.00000$ \\
\shortstack{Post-fault\\recovery shaping} & Step-like & Limited & Ramp-limited \\
\hline\hline
\end{tabularx}
\end{table}

\begin{table}[!htbp]
\caption{Stress Tests (Constraint Binding and Graceful Degradation; fault-window metrics)}
\vspace{-0.6\baselineskip}
\label{tab:stress}
\centering
\scriptsize
\setlength{\tabcolsep}{2.5pt}
\renewcommand{\arraystretch}{1.05}
\begin{tabularx}{\columnwidth}{@{}>{\raggedright\arraybackslash}Xccccc@{}}
\hline\hline
\textbf{Case} & \shortstack{$\min|V_{pcc}|$\\(during dip)} & $\max|I|$ & $\overline{P}_{draw}$ & $\min V_{dc}$ & \shortstack{Unserved\\(MWh)} \\
\hline
Baseline & 0.482 & 0.570 & 0.198 & 0.784 & 0.00000 \\
Low DC energy ($T_{dc}=0.1$ s) & 0.492 & 0.866 & -0.157 & 0.699 & 0.00071 \\
Detection delay ($t_{det}=10$ ms) & 0.478 & 0.569 & 0.192 & 0.784 & 0.00000 \\
Slow BESS response ($\tau_{bess}=0.10$ s) & 0.493 & 0.799 & 0.140 & 0.746 & 0.00009 \\
Low BESS power ($P_{dis}^{max}=0.3$) & 0.019 & 0.787 & 0.043 & 0.752 & 0.00137 \\
Low BESS ramp ($R_{bess}=1.0$) & 0.443 & 0.797 & 0.024 & 0.700 & 0.00104 \\
No BESS ($P_{dis}^{max}=0$) & 0.037 & 0.999 & 0.042 & 0.700 & 0.00200 \\
\hline\hline
\end{tabularx}
\end{table}

\begin{table}[!htbp]
\caption{Mode 2 Ablation Summary (Fault-Window Metrics)}
\vspace{-0.6\baselineskip}
\label{tab:ablation}
\centering
\scriptsize
\setlength{\tabcolsep}{2.5pt}
\renewcommand{\arraystretch}{1.05}
\begin{tabularx}{\columnwidth}{@{}>{\raggedright\arraybackslash}Xccccc@{}}
\hline\hline
\textbf{Variant} & \shortstack{$\min|V_{pcc}|$\\(during dip)} & $\max|I|$ & $\overline{P}_{draw}$ & $\min V_{dc}$ & $\max|\dot{P}_{draw,ref}|$ \\
\hline
Baseline & 0.482 & 0.570 & 0.198 & 0.784 & 0.200 \\
No Q-support ($K_v=0$) & 0.396 & 0.354 & 0.198 & 0.784 & 0.200 \\
No min draw policy ($P_{draw,min}^{fault}=0$) & 0.482 & 0.517 & 0.095 & 0.784 & 0.200 \\
No soft return & 0.037 & 0.993 & 0.042 & 0.811 & 74.101 \\
No BESS & 0.037 & 0.999 & 0.042 & 0.700 & 0.200 \\
\hline\hline
\end{tabularx}
\end{table}

\noindent\textit{Note on very low $\min|V_{pcc}|$ values:} In constraint-binding cases (e.g., no UPS-BESS buffering or step-like recovery), the averaged model can collapse under the imposed limits; extremely low instantaneous $\min|V_{pcc}|$ indicates infeasible ride-through. We therefore also report the settled $\min|V_{pcc}|$ metric (Table~\ref{tab:comparison}).

\subsubsection{Robustness Sweeps (Balanced Dip)}
We run small one-at-a-time sweeps on grid strength and retained fault voltage magnitude $V_{dip}$ using the same Mode~2 controller and limits. Table~\ref{tab:sweeps} reports the settled minimum PCC voltage (after one 60~Hz cycle), peak inverter current magnitude, peak UPS-BESS power magnitude during the fault, and the imposed ramp limit $\max|\dot{P}_{draw,ref}|$.

\begin{table}[!htbp]
\caption{Robustness Sweeps (Proposed Controller; Balanced Three-Phase Dip, Fault-Window Metrics). Here, $V_{dip}$ denotes the retained PCC-voltage magnitude applied in the Thevenin source during the fault window.}
\vspace{-0.6\baselineskip}
\label{tab:sweeps}
\centering
\scriptsize
\setlength{\tabcolsep}{2.5pt}
\renewcommand{\arraystretch}{1.05}
\begin{tabularx}{\columnwidth}{@{}>{\raggedright\arraybackslash}Xcccc@{}}
\hline\hline
\textbf{Sweep} & \shortstack{$\min|V_{pcc}|$\\(after 1 cycle) (p.u.)} & \shortstack{$\max|I|$\\(p.u.)} & \shortstack{$\max|P_{bess}|$\\(p.u.)} & \shortstack{$\max|\dot{P}_{draw,ref}|$\\(p.u./s)} \\
\hline
SCR=1.2 ($V_{dip}=0.5$) & 0.811 & 0.513 & 0.924 & 0.200 \\
SCR=1.5 ($V_{dip}=0.5$) & 0.789 & 0.570 & 0.924 & 0.200 \\
SCR=2.0 ($V_{dip}=0.5$) & 0.757 & 0.654 & 0.924 & 0.200 \\
SCR=3.0 ($V_{dip}=0.5$) & 0.711 & 0.781 & 0.924 & 0.200 \\
\hline
$V_{dip}=0.4$ (SCR=1.5) & 0.740 & 0.687 & 0.924 & 0.200 \\
$V_{dip}=0.5$ (SCR=1.5) & 0.789 & 0.570 & 0.924 & 0.200 \\
$V_{dip}=0.7$ (SCR=1.5) & 0.873 & 0.364 & 0.924 & 0.200 \\
\hline\hline
\end{tabularx}
\end{table}

\noindent Across these sweeps, unserved IT energy remains effectively zero under the modeled UPS-BESS limits. As expected, increasing $P_{draw,min}^{fault}$ increases active current retention at the expense of reactive voltage support, while increasing $R_{limit}$ proportionally increases $\max|\dot{P}_{draw,ref}|$ and shifts more of the post-fault recovery burden onto the UPS-BESS.

\subsubsection{EMT-Style Time-Step Validation (Non-Switching, $abc$ Domain)}
\begin{figure}[htbp]
\centerline{\includegraphics[width=\columnwidth]{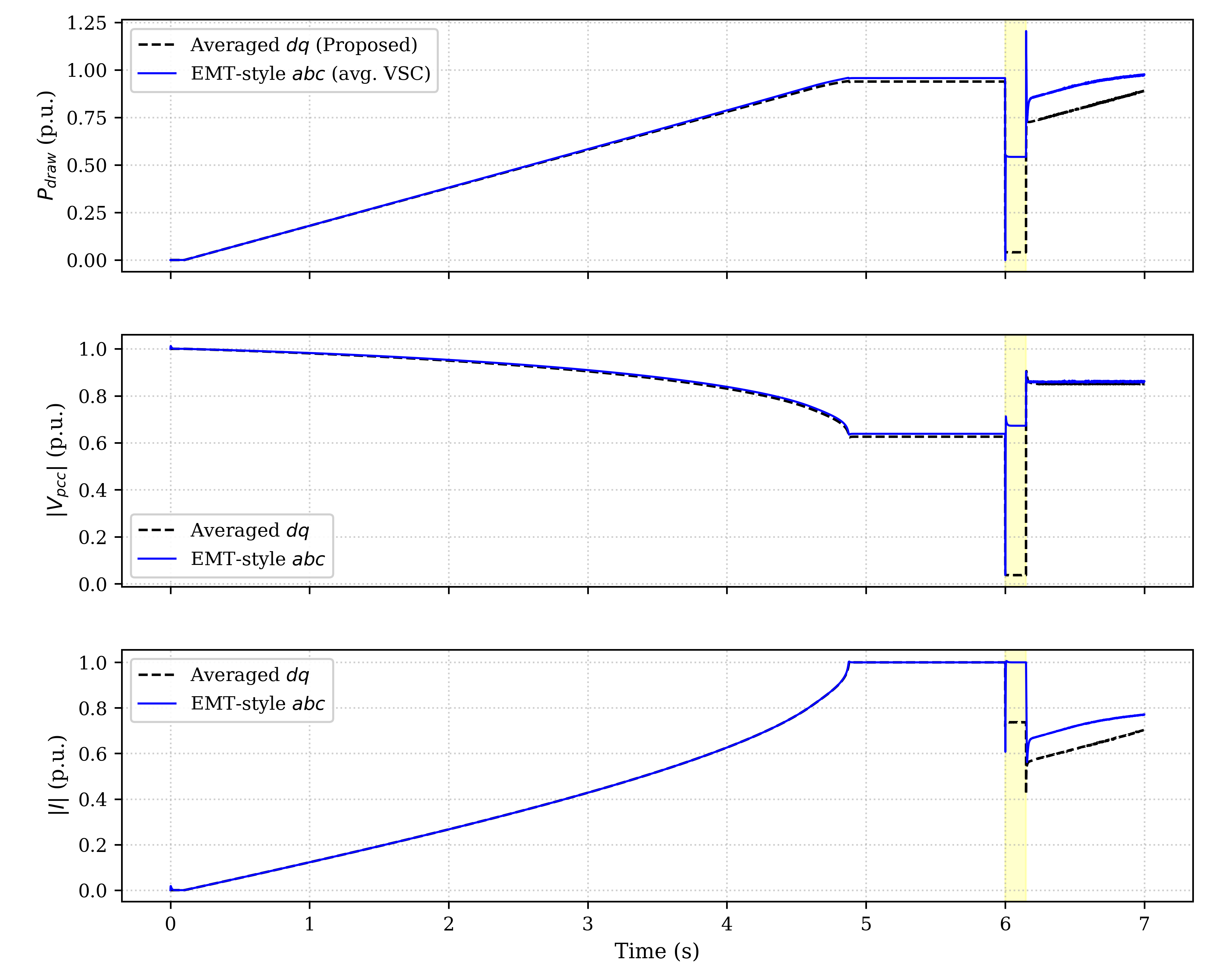}}
\caption{Comparison of averaged-$dq$ and EMT-style fixed-step $abc$ simulation for the proposed controller (non-switching, commanded-voltage VSC). To allow a settled pre-fault operating point in both models, the voltage dip is applied later in the window (yellow band, $t=6.0$--$6.15$~s). For apples-to-apples comparison, the EMT traces are post-processed into the same phasor-equivalent Thevenin PCC quantities used in the averaged model. Panels show (top) $P_{draw}$, (middle) $|V_{pcc}|$, (bottom) $|I|$. While startup and recovery match, the fault window reveals that the averaged model is conservative, predicting voltage collapse under the quasi-static algebraic grid constraint. In contrast, the EMT dynamics (specifically grid inductance) physically constrain the rate of voltage decay, allowing the fast controller to intervene and maintain stability. This confirms the averaged model as a safe, lower-bound design tool.}
\label{fig:emt_compare}
\end{figure}

Unbalanced/negative-sequence behavior is outside the scope of the balanced averaged-$dq$ model and is left for EMT and hardware-in-the-loop (HIL) validation. As an intermediate check, we implement a fixed-step, non-switching $abc$ simulation of the same control logic ($\Delta t=50~\mu$s) with a commanded voltage-source behind an $RL$ filter connected to an $RL$ Thevenin grid (derived from the same SCR/X/R). Fig.~\ref{fig:emt_compare} compares the response under full-load conditions ($t=6.0$~s). While startup and recovery match, the fault window reveals that the averaged model is conservative, predicting voltage collapse under the quasi-static algebraic grid constraint. In contrast, the EMT dynamics (specifically grid inductance) physically constrain the rate of voltage decay, allowing the fast controller to intervene and maintain stability. This confirms the averaged model as a safe, lower-bound design tool.

\subsection{Mode 3: Frequency Support via Grid-Draw Modulation}
\subsubsection{Bulk-System Frequency Proxy and Closed-Loop Demonstration}
\begin{figure}[htbp]
\centerline{\includegraphics[width=\columnwidth]{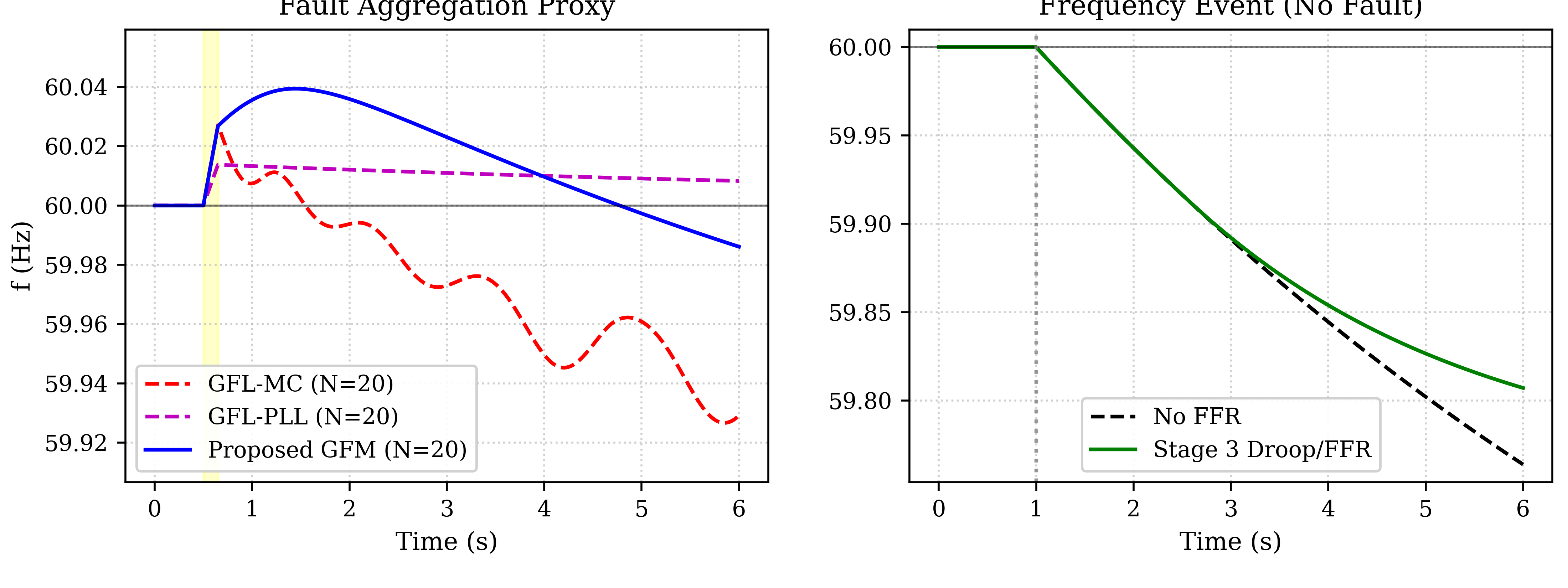}}
\caption{Bulk-frequency proxy results: (left) fault aggregation mapping comparing GFL-MC (red), GFL-PLL (magenta), and proposed controller (blue); (right) closed-loop no-fault frequency event co-simulation showing Mode~3 droop/FFR via grid-draw modulation.}
\label{fig:freq_proxy}
\end{figure}

To connect the fault-mode load behavior to bulk-system impacts and provide a concrete Mode~3 droop/FFR controller example, we couple the aggregated data-center model to a simple swing-equation frequency proxy on an external system base. The proxy models the aggregated power imbalance as:
\begin{equation}
2H\frac{dx}{dt} = \left(P_m - P_e\right) - D x
\end{equation}
where $x=(f-f_0)/f_0$, $P_e$ is the aggregated data-center draw mapped to a bulk system base, and $P_m$ is fixed at the pre-event operating point. Fig.~\ref{fig:freq_proxy} (left) maps the fault-case $P_{draw}$ trajectories for an aggregation of $N=20$ identical 50~MW blocks on a 30~GW system base, illustrating that avoiding momentary cessation and enforcing a minimum fault-mode draw reduces the associated frequency excursion. The proxy is order-of-magnitude only; more detailed analyses are available in \cite{tayyebi2020frequency, kenyon2021interactive}.

Fig.~\ref{fig:freq_proxy} (right) shows a no-fault frequency-event co-simulation with Mode~3 droop/FFR acting on $P_{draw,target}$ (Section~\ref{sec:stage3}). The controller computes $P_{supp}^{sys}=\mathrm{sat}(K_f(-x),\pm P_{FFR,max}^{sys})$, maps it to the data-center base, and modulates $P_{draw,target}$ using the same shaping filter and UPS-BESS limits as Mode~2; Mode~2 disables Mode~3 during fault/low-voltage operation.

\textbf{Limitations/sensitivity:} This is not a full interconnection frequency model (no governor, underfrequency/undervoltage load shedding (UFLS/UVLS), or network electromechanical modes). As expected from the swing equation, the deviation scales roughly with $N/H$; in a small sweep over $N\in\{10,20,40\}$ and $H\in\{3,5,7\}$~s (fixed $D$), the worst case ($N=40$, $H=3$~s) reaches $\approx 60.159$~Hz for GFL-MC and $\approx 59.953$--$59.961$~Hz for the proposed controller.

%% file: sections/05_conclusion.tex
\section{Conclusion}
\label{sec:conclusion}

This paper proposed a three-mode supervisory control framework for centralized MV-UPS systems: Mode~1 regulates a DC stiff bus for internal continuity, Mode~2 provides grid-friendly fault resilience (current-limited reactive priority with a minimum-draw policy) and rate-limited post-fault ``soft return,'' and Mode~3 optionally provides fast frequency response via grid-draw modulation.
Averaged-$dq$ simulations at an ultra-weak node (SCR~=~1.5) show reduced peak inverter current, improved sustained PCC voltage during a balanced three-phase dip, elimination of unserved IT energy within modeled UPS-BESS limits, and smoother post-fault recovery versus grid-following benchmarks.
In addition, normal-operation studies show (i) DC stiff-bus error rejection from an initial DC-link energy deficit and (ii) supervisory power shaping that attenuates a 1~Hz pulsed load component as seen by the grid, while the UPS-BESS absorbs/supplies the oscillatory component within its constraints.
Future work will focus on EMT/HIL validation, unbalanced-fault and protection interactions, and coordination across multiple facilities.

%% file: sections/05b_ai_disclaimer.tex
\begingroup
\scriptsize
\noindent\textit{AI-use statement:} During preparation of this work the author used Gemini~3.0 Pro to improve language clarity; the author reviewed/edited the manuscript and takes responsibility for all content.
\endgroup